\def\Version{{6}} 





\message{<< Assuming 8.5" x 11" paper >>}    

\magnification=\magstep1	          
\raggedbottom

\parskip=9pt

%

\def\singlespace{\baselineskip=12pt}      
\def\sesquispace{\baselineskip=16pt}      


\input epsf
\epsfverbosetrue






\font\openface=msbm10 at10pt
 %

\def\Reals         {{\hbox{\openface R}}}
\def\Complexes     {{\hbox{\openface C}}}

 %
 %
 %






\def\Re  {\mathop{\rm Re}  \nolimits}    
\def\Im  {\mathop{\rm Im}  \nolimits}    
\def\Imag  {\mathop{\rm Imag}  \nolimits}    





%

%
%



%



\def\sqr#1#2{\vcenter{
  \hrule height.#2pt 
  \hbox{\vrule width.#2pt height#1pt 
        \kern#1pt 
        \vrule width.#2pt}
  \hrule height.#2pt}}




\def\lto{\mathop
        {\hbox{${\lower3.8pt\hbox{$<$}}\atop{\raise0.2pt\hbox{$\sim$}}$}}}
\def\gto{\mathop
        {\hbox{${\lower3.8pt\hbox{$>$}}\atop{\raise0.2pt\hbox{$\sim$}}$}}}
%
%
%



\def\part{\subseteq}		

\def\braces#1{ \{ #1 \} }



\def\to{\mathop\rightarrow}	




\def\interior #1 {  \buildrel\circ\over  #1}     




\def\basisvector#1#2#3{
 \lower6pt\hbox{
  ${\buildrel{\displaystyle #1}\over{\scriptscriptstyle(#2)}}$}^#3}



\def\hat{\widehat}		





\fontdimen16\textfont2=2.5pt
\fontdimen17\textfont2=2.5pt
\fontdimen14\textfont2=4.5pt
\fontdimen13\textfont2=4.5pt




%

%
 \let\miguu=\footnote
 \def\footnote#1#2{{$\,$\parindent=9pt\baselineskip=13pt%
 \miguu{#1}{#2\vskip -7truept}}}
%
%

\def\linebreak{\hfil\break}
\def\lbr{\linebreak}
\def\pagebreak{\vfil\break}


\def\BulletItem #1 {\item{$\bullet$}{#1 }}
\def\bulletitem #1 {\BulletItem{#1}}

\def\PrintVersionNumber{
 \vskip -1 true in \medskip 
 \rightline{version \Version} 
 \vskip 0.3 true in \bigskip \bigskip}

\def\author#1 {\medskip\centerline{\it #1}\bigskip}

\def\address#1{\centerline{\it #1}\smallskip}

\def\furtheraddress#1{\centerline{\it and}\smallskip\centerline{\it #1}\smallskip}

\def\email#1{\smallskip\centerline{\it address for email: #1}} 

\def\AbstractBegins
{
 \singlespace                                        
 \bigskip\leftskip=1.5truecm\rightskip=1.5truecm     
 \centerline{\bf Abstract}
 \smallskip
 \noindent	
 } 
\def\AbstractEnds
{
 \bigskip\leftskip=0truecm\rightskip=0truecm       
 }

\def\section #1 {\bigskip\noindent{\headingfont #1 }\par\nobreak\noindent}

\def\subsection #1 {\medskip\noindent{\subheadfont #1 }\par\nobreak\noindent}
 %

\def\ReferencesBegin
{
 \singlespace					   
 \vskip 0.5truein
 \centerline           {\bf References}
 \par\nobreak
 \medskip
 \noindent
 \parindent=2pt
 \parskip=6pt			
 }
 %

\def\reference{\hangindent=1pc\hangafter=1} 

\def\ref{\reference}

\def\sepref{\parskip=4pt \par \hangindent=1pc\hangafter=0}
 %

\def\journaldata#1#2#3#4{{\it #1\/}\phantom{--}{\bf #2$\,$:} $\!$#3 (#4)}
 %

\def\eprint#1{{\tt #1}}

 %

\def\webtilde{\lower2pt\hbox{${\widetilde{\phantom{m}}}$}}

\def\webhome{{\tt {http://www.physics.syr.edu/}{\webtilde}{sorkin/}}}
 %

 %


\def\hpf#1{\webhome{\tt{some.papers/}}}
 %

\def\hpfll#1{\webhome{\tt{lisp.library/}}}
 %


\font\titlefont=cmb10 scaled\magstep2 

\font\headingfont=cmb10 at 12pt
%

\font\subheadfont=cmssi10 scaled\magstep1 
%







\def\xdot{\dot{x}}
\def\tdot{\dot{t}}



\phantom{}


\PrintVersionNumber



\sesquispace
\centerline{{\titlefont Is the spacetime metric Euclidean rather than
 Lorentzian?}\footnote{$^{^{\displaystyle\star}}$}%
%
{published in {\it Recent Research in Quantum Gravity}, edited by A. Dasgupta (Nova Science Publishers NY, 2013)
 \eprint{arxiv 0911.1479 [gr-qc]}
}}

\bigskip


\singlespace			        

\author{Rafael D. Sorkin}
\address
 {Perimeter Institute, 31 Caroline Street North, Waterloo ON, N2L 2Y5 Canada}
\furtheraddress
 {Department of Physics, Syracuse University, Syracuse, NY 13244-1130, U.S.A.}
\email{sorkin@physics.syr.edu}

\AbstractBegins                              
   My answer to the question in the title is ``No''.  In support of this
   point of view, we analyze some examples of saddle-point methods,
   especially as applied to quantum ``tunneling'' in
   nonrelativistic particle mechanics and in cosmology.  Along the way we
   explore some of the interrelationships among different ways of
   thinking about path-integrals and saddle-point approximations to
   them. 
\AbstractEnds

\noindent
\centerline{\bf Contents}
\smallskip
\centerline{\it The saddle-point method as applied to a simple integral}
\centerline{\it One dimensional tunneling}
\centerline{\it Tunneling in quantum cosmology}
\centerline{\it Summary}



\sesquispace



\section{}			
The use of non-Lorentzian metrics (complex metrics or metrics of
Euclidean signature) as instantons or in connection with black hole
thermodynamics has sometimes provoked the opinion that spacetime is
``really'' a Riemannian geometry as opposed to a Lorentzian one; and
given this, it may not be out of place to adduce here certain more or
less obvious points to the contrary.
In the following I will illustrate these points with some simple
and --- I hope --- instructive
examples of saddle-point approximations of the type that arise in
``tunneling'' processes, including the process of the birth of a cosmos,
as it has been studied in ``quantum cosmology''.
Much of what I will say is well known, but perhaps some of it is not as
familiar as it might be, and I hope that the examples can help to
clarify certain confusing aspects that arise in connection with
saddle-point methods and the analytic continuation of path integrals.
Perhaps also, the cosmology example can serve to bring out the
conceptual (and potentially technical) advantage to be gained in
quantum gravity by unambiguously taking all amplitudes to be defined
in the first instance by Lorentzian path integrals.

Of course, it would be rash to reject dogmatically the possibility that
quantum gravity might ultimately teach us to think of spacetime as more
like a Riemannian geometry than a Lorentzian one, but it is fair, I
think, to maintain that the usefulness of complex metrics as instantons
provides no evidence on the question one way or the other.  Rather, it
seems much more straightforward and natural to interpret complex
spacetimes simply as 
contributors to
an analytically continued path
integral that originally is taken over a space of Lorentzian metrics.
If that interpretation is correct then for example, an amplitude
computed from a Euclidean-signature instanton should be thought of as
compounded from
the contributions of a large number of Lorentzian histories, 
and the instanton itself should be thought of as merely a saddle-point
of the analytically continued integrand.

\section{The saddle-point method as applied to a simple integral}
Below, I will illustrate what I mean, using the example of tunneling in
nonrelativistic quantum mechanics, but to lay the groundwork for that
discussion, it seems appropriate to recall in some detail how the saddle
point method works in the one-dimensional case, 
as expounded in textbooks covering complex analysis [1].  
This method is designed for integrals of the general form
$$
            \int g(x) \, dx \, \exp(f(x))  \ ,  \eqno(1)
$$
where the exponent $f(x)$ is supposed to be ``rapidly varying'' and 
$g(x)$ (the ``measure-factor'' in path-integral argot) is supposed to be
``slowly varying with respect to $f(x)$''.
An instructive instance of this general form for us is
$$
    I(A) = \int_0^\infty {dx\over x^{3/2}} \exp\braces{A(ix-1/x)}  \ ,
   \eqno(2)
$$
where $A$ is an arbitrary positive parameter corresponding to $1/\hbar$
in the path-integral.
We have then $f(x)=A(ix-1/x)$ and $g(x)=x^{-3/2}$.
In order to satisfy the condition of ``rapid variation'', I will assume
further that $A\gg1$.  Not only does this integral bear a strong
resemblance to a path-integral, but it can be evaluated exactly as
$$
    I(A) = \sqrt{\pi \over A} \exp(\sqrt{2} \, A (i-1)) \ ,  \eqno(3)
$$
thanks to our specific choice of $g(x)$.\footnote{$^\dagger$}
{This, and not any specific analogy with a path-integral, was
 the reason why I chose $3/2$ as the power of $x$.  Incidentally, the 
 resemblance to a path-integral would have been even closer had we 
 taken the exponent to be $iA(x-1/x)$, but then (2) would
 have failed to be absolutely convergent.}

%

The integrand in (2) analytically continues to the entire complex
plane $\Complexes$ except for $z=0$, where it has an essential singularity.  
Its saddle points are by definition the solutions of $f'(z)=0$,
namely
$z_\pm=\pm\sqrt{i}=\pm(1+i)/\sqrt{2}$, 
and at these points we have
$f(z)=2izA$,
$f''(z)=2Az$, 
$f'''(z)=-6A$.  
%
%
As a rule of thumb, a saddle-point approximation is valid
when $|f'''(z)|^2\ll|f''(z)|^3$ at the saddle-point,
and this holds in consequence of our assumption that $A\gg1$.

Since there are two saddle points, the question arises which if either
we should use, but let's leave that aside for a moment.
To each saddle point belongs a potential contribution to the integral,
given up to sign by the general formula
$$
        g(z) \  \sqrt{2\pi \over - f''(z)} \   e^{f(z)}    \ , \eqno(4)
$$
or in our case,
$$
        {1\over z^{3/2}} \  \sqrt{2\pi \over - f''(z)} \   e^{f(z)} \ . \eqno(5)
$$
Remarkably, the saddle point at $z=+\sqrt{i}$ yields the exact answer\footnote{$^\flat$}
{This accidental exactness would presumably not have happened had we chosen to
 absorb the prefactor $x^{-3/2}$ into the exponent $f(z)$.}
if we adopt the correct sign for the square root in 
(5).  But how do we know which
sign to choose without knowing the answer in advance, and how do we know
which of the saddle points actually contributes?
Both questions can be decided by ad hoc considerations in this case, but
only if we already know that a saddle point approximation is actually
valid.  Concerning the question of which saddle point, one can notice
first of all that $I(A)\to0$ as $A\to\infty$, and compare this with the
leading behavior of the two respective exponentials, namely
$\exp(f(z\pm))=\exp(\pm 2A i^{3/2})$.  Since $\exp(-2A i^{3/2})$ blows
up with $A$, only the saddle point at $z=z_+=\sqrt{i}$ can be correct.
The question of the sign can be decided by observing that $I(A)$ becomes
a positive number as $A\to0$, but this is even more of a cheat since
$A\ll1$ is altogether outside the domain of validity of the saddle
point approximation. 
(Notice in this example that the correct saddle point is the one at
which the integrand is smaller, not larger.)

Now let us approach the question more systematically.  We started with
an integral whose path or ``contour'' $\Gamma=\Gamma_0$ was the positive
real axis, traversed from $0$ to $\infty$.  
We then analytically
continued the integrand and found two complex saddle points $z_\pm$.  
In order to make use of a saddle point, however, we need to
deform $\Gamma$ to pass through it, and it is necessary that this
deformation not alter the value of the integral.  In particular this
means that $\Gamma$ should not cross a singularity of the integrand (or
if it does the extra contribution should be evaluated).  In our case,
this only has the effect of preventing us from rotating $\Gamma$ freely
about its initial endpoint, i.e. it requires the deformed contour to
depart from the origin in the direction of positive $x$.  To this
extent, both saddle points are accessible.

However, the next condition is that, along the deformed contour
$\Gamma$, the biggest contribution to the integral arise in the
neighborhood of the saddle point, and not somewhere else.  Normally, one
tries to ensure this by arranging that away from the saddle point,
either the integrand dies out rapidly (steepest descent method) or it
suffers cancellation from rapid variations in its phase (stationary
phase method).  For the former method, the most favorable path through
the saddle point is that of ``steepest descent'', along which the
imaginary part of $f(z)$ remains constant.  Along this path the phase of
$\exp(f(z))$ remains constant, while its magnitude drops off rapidly as
one departs from the saddle point.  Given that $f''(z)\not=0$ at the
saddle point, there will be exactly one such path passing through it.
In our case, and for the saddle point $z_+$, this path, a solution of
the equation $\Im f(z) = constant$, or
$$
         x + {y \over x^2 + y^2} = constant =  \sqrt{2} \ , \eqno(6)
$$
originates at $z=0$ with zero slope, curves upward passing through
$z=z_{+}$, and goes off to infinity in the direction parallel to the
positive imaginary axis.  
Clearly it is possible to deform such a path to our
original contour $\Gamma_0$ by adjoining a large circular arc
near infinity, and this won't affect our integral since our integrand
drops off rapidly there.  (See Figure 1.)
Hence, the saddle point $z=z_{+}$ is
indeed ``accessible'', and we obtain the approximation (5) to $I(A)$.
Finally, the unknown sign in (5) is determined by the rule
that the phase of the square root factor should coincide with that of
the tangent direction to the steepest descent contour $\Gamma$ at the
saddle point.  The latter is easily evaluated and leads to the answer
(3).

We could also have obtained this answer using a stationary-phase contour
through $z=z_{+}$.  Along such a curve, it is the real part of $f(z)$
that remains fixed, leading to an integrand of varying phase and
(almost) constant magnitude.\footnote{$^\star$}
{It would have been exactly constant if not for the prefactor $g(z)$ in
 the integrand.  Through any saddle point (meaning a point where $f'$
 vanishes, but not $f''$), there will be two paths of constant
 $\Re{f(z)}$ meeting at right angles and rotated by 45 degrees from the
 steepest descent path.  This makes a total of three rays in the tangent
 space.  The fourth ray needed to complete the pattern (the one
 orthogonal to the steepest descent ray) might be called the direction
 of steepest {\it ascent}.  It represents in some sense the worst
 possible choice for an integration contour.}
In our case, there is such a contour that rises vertically from $z=0$,
bends over to pass through $z_{+}$, and continues on to $\infty$,
moving parallel to the $x$-axis.  This also is 
a valid 
deformation of our
original contour and leads to the same result (3).  
(For a stationary-phase path, one can determine the sign by applying the
aforementioned rule to the ``adjacent'' steepest descent direction.)

Now what about the other saddle point $z_{-}$?  For consistency it ought
{\it not} to be accessible, since if it were, we would have two
disagreeing approximations to the same integral.  To see what goes
wrong consider first the possible steepest descent contours through
$z=z_{-}$.  There is a curve of constant $\Im{f(z)}$ symmetric to the
steepest descent curve through $z=z_{+}$, but it actually is a curve of
``steepest ascent'', as shows up clearly in Figure 1.  The true steepest
descent contour is the other curve of constant $\Im{f(z)}$ through
$z=z_{+}$.  It emerges from the origin horizontally to the right, bends
downward and around to pass through $z=z_{-}$, and then continues upward
to infinity parallel to the imaginary axis.  By adding a circular
arc at infinity as before, we could draw this curve down to the
positive real axis, but 
the resulting contour
would not be deformable to our original
contour $\Gamma_0$, since it now would wind once around the origin.  
In order to unwind it, we would have to 
transport its ``endpoint at infinity'' 
in the opposite (counterclockwise) direction to what we just
considered, 
but then the added circle at infinity would make an
exponentially great contribution to the integral, as one sees in Figure 1.
(Instead of adding an arc to the steepest descent contour near infinity,
we could deviate from it at finite radii, but as Figure 1 shows, we would
still have to pass through a region where the integrand would be
exponentially greater than in the neighborhood of the saddle point.)
We might also consider using one of the stationary-phase contours
through $z=z_{-}$,
but we would run into essentially the same obstruction with them as
well. 
Either the contour would wind around the origin or the main
contribution would fail to (be guaranteed to) come from the neighborhood
of the saddle point.

I probably have lavished on this simple calculation more explanation
than it deserves, but I wanted to illustrate that the mere existence of
a saddle point is no guarantee of a corresponding saddle-point
approximation.  Rather, as we have seen, several conditions need to be
satisfied which allow one to deform the original integration contour
into one passing through the saddle point (or points) in such a manner
as to validate the approximation (4).  The cases of functional-
or path-integrals are much more complicated, but it should be borne in
mind that analogous conditions apply to them as well.  As far as I know,
these conditions are generally ignored entirely in the case of gravity,
and people have never even tried to formulate them in that case.

In this connection, let us have a look at a different technique of
analytic continuation which can also shed light on the  
quantum mechanical and gravitational cases, 
and which, 
when it applies, 
is in some ways simpler.  
In our integral (2), rather than analytically
continuing the function $f$ in its argument $x$, we can contemplate
instead continuing in the parameter $A$.  Actually, it's more convenient to 
rescale $x$ or equivalently to introduce a second parameter that
multiplies only the first term in $f$.   Our integral then takes the
form $I(A)=J(B)B^{-1/4}$, where $B=A^2$ and
$$
      J(B)=  \int_0^\infty {dx\over x^{3/2}} \exp(iBx-1/x)  \ .
$$
If now we analytically continue $B$ to positive imaginary values, 
the integrand will become purely real
and will acquire a saddle-point.
In order that $J(B)$ be holomorphic in $B$
the integral must continue to exist, 
however,
and in order that the integral
continue to exist, we need to continue $B$ through the upper half-plane.
This fixes the direction of the continuation.  
When $B=ib$ with $b$ real and positive, 
the exponent, $-(bx+1/x)$,
has a single saddle point at $x=1/\sqrt{b}$, and its value there is
simply $-2\sqrt{b}$ or in other words $-2\sqrt{-iB}$.
(The point $x=-1/\sqrt{b}$ needn't be considered at all since it is
 outside the range of integration.)
It is now easy to evaluate 
the contribution of this saddle-point to $J(B)$, 
and there is no longer any question of a sign ambiguity.  
The only small subtlety arises
when we analytically continue the answer,
$$
         J(B) = \sqrt{\pi} \exp(-2\sqrt{-iB})  \ ,
$$
back from imaginary to real $B$ (a well-defined procedure because we
know $J(B)$ on the entire positive imaginary axis).  
Being careful to move in a clockwise
direction so that $B$ remains in the upper half-plane, we find that
the exponent becomes $+\,i^{3/2}2\sqrt{B}$ = $\sqrt{2}A(i-1)$,
and we recover our previous answer (3).\footnote{$^\dagger$}
{Here, I have ignored a second subtlety, perhaps less small.  How do we
 know that our approximate answer will still be approximately correct
 after we have continued it back to real $B$?}

\section{One dimensional tunneling}
With the above example in mind, let us turn now to the case of
one-dimensional tunneling through a potential barrier.  
Imagine, then, a (non-relativistic)
particle emitted at $x=x_0$
by a ``source of energy $E$'' 
and 
absorbed by a ``sink of energy $E$'' at
$x=x_1$.  
The amplitude for this process is 
formally 
$$
       \int\limits_{\gamma\in\Gamma} d\gamma \; \exp\braces{{i\over \hbar} \, S(\gamma)}  \ ,
   \eqno(7)
$$
where
$$
   S(\gamma) = \int\limits_\gamma {m\over 2} {dx^2\over dt} - V(x) dt \ + \ E(t_1 - t_0) \ .
   \eqno(8)
$$
Here $t_0$ and $t_1$ are respectively the times of emission and absorption,
and 
the infinite-dimensional ``integration contour'' 
$\Gamma=\Gamma(x_0,x_1)$ 
comprises all trajectories $\gamma$ from $x_0$ to $x_1$ 
($t_0$ and $t_1$ being left free).
When we come to analytically continue in the trajectory $\gamma$, it
will be useful to have expressed $S(\gamma)$ also in so-called
parameterized form,\footnote{$^\flat$}
{The distinction between the ``parameterized'' and ``unparameterized''
 forms of the path-integral is not an independent question of it own,
 but an aspect of the question of ``choice of measure-factor''.}
i.e. as a worldline in $\Reals\times\Reals$, coordinatized by an
arbitrary parameter $\lambda$ that we can take to run from $0$ to $1$.
So expressed, $S$ takes the form
$$
  S = \int_0^1 d\lambda 
   \left( {m \xdot^2 \over 2 \tdot} +  (E-V(x)) \, \tdot  \right)  \ .
  \eqno(9)
$$
Now the integral (7) is not only formal, but even formally it is
infinite 
because an infinite amplitude accumulates in an infinite time.  In a
more complete treatment, we would have to deal with this problem of
normalization, but here my only interest is in the role played by
the complexification of $t$.  I will therefore ignore this issue
together with all other technical issues that seem irrelevant to the
main question (including for example the question of the ``measure
factor'' hidden within $d\gamma$,  the analog of $g(x)$ in our
earlier example).

Now as usual, the choice of $E$ divides $\Reals$ into ``classically
allowed'' regions where $E-V(x)>0$ and ``classically forbidden'' regions
where $E-V(x)<0$.  Let us assume to begin with that both $x_0$ and $x_1$
are either within or on the edge of the forbidden region (which we
assume to be connected).
If instead, $[x_0,x_1]$ lay in the allowed region
then a classical trajectory $\gamma_0$ would join $x_0$ to $x_1$ and
(such a trajectory being synonymous with a saddle-point of $e^{iS(\gamma)}$) 
we could then 
--- without in any way modifying our ``integration contour'' $\Gamma$ --- 
directly apply the method of stationary phase to (7),
assuming $\hbar$ to be sufficiently small.
This is the familiar procedure involved in deducing a classical limit
from the path integral.

It bears emphasis here that although we cannot take a similar classical
limit within the forbidden region, there's nothing wrong in the
slightest with trajectories that traverse that region.
The path integral (7) makes equal sense for all paths $\gamma$ and
for all values of the energy.  What is special about having $x_0$ and
$x_1$ in the forbidden region is only that we can no longer find a
saddle point within the corresponding domain of integration 
$\Gamma(x_0,x_1)$.
To obtain a useful approximation to  
(7) we therefore have either to find another method or
to resort to an analytic continuation.  
As before, this can be approached in more than one way.  We can
``bring the saddle point to $\Gamma$'',
as we did above when we continued $A$ into the complexes; 
or we can ``bring $\Gamma$ to the saddle point'',
as we did above when we continued $z$ into the complexes 
and then deformed $\Gamma$.
Let us pursue the former avenue first, 
since it seems conceptually simpler than the latter.

Defining $W(x)=V(x)-E$, 
we can write (9) more compactly as
$$
   S(\gamma) = \int\limits_\gamma {m\over 2} {dx^2\over dt} - W(x) dt \ ,
   \eqno(10)
$$
where of course  
$dx$ and $dt$ are increments of $x$ and $t$ 
along the worldline $\gamma$.
Under what circumstances will we be able to find a saddle point
within our domain of integration $\Gamma(x_0,x_1)$?
The condition for $\gamma\in\Gamma$ to be a saddle point is that it be a
critical point of $S(\gamma)$, in other words $S$ must be unchanged by
small variations of $\gamma$.
Since the limits $t_i$ are not fixed, the most general variation of
$\gamma$ involves not only varying $x$ at fixed $t$ but also varying the
range of $t$.
The latter is equivalent to varying $dt$ freely in (10) 
[or $\tdot$ freely in (9)],
and it produces immediately the condition,
$$
    {m\over 2} \left({dx\over dt}\right)^2 + W(x) = 0  \ .  \eqno(11)
$$
Given this ``energy equation'', the variation of $x$ can be ignored, since it
adds almost nothing to what one obtains by differentiating (11).

Within the forbidden region, $W$ is positive, and (11)
obviously has no solution.
To remedy this lack, it suffices to change
the sign either of $W(x)$ or of $m$; but the latter looks more promising
since $W$ is a function while $m$ is only a number.  Accepting this
hint, let us regard $S$ as a function
of both $\gamma$ and $m$,
and analytically
continue it from positive to negative values of $m$.  In so doing, we
must however be careful that our amplitude (7) [call it $Z(m)$]
remain holomorphic throughout the region through which $m$ will pass.
Otherwise it will not be possible to analytically continue 
$Z$
back to positive $m$.  But it is easy to see what this requires, 
given that, 
insofar as the term in (9) involving $m$ is concerned, 
(7) is just a Gaussian integral.  
As long as $\Im{m}>0$ the Gaussian will look like $e^Q$ with the real
part of $Q$ being a negative-definite quadratic form in $\xdot$.
Therefore, the Gaussian will decay at
infinity and the integral defining $Z(m)$ will exist and be holomorphic in $m$.  
However $m$ must avoid the
lower half-plane because 
the real part of $Q$ is positive there,
and the integral would diverge.  
Thus, holomorphicity requires that $m$ circle the origin in a
counterclockwise sense.

When $m$ is negative, 
(11) does have a solution $\gamma$,
and plugging this back into (10) 
(or using as a shortcut the general fact that the expression $A/x+Bx$
assumes the value $\sqrt{AB}$ at its extremum),
it is then easy to compute the resulting,
stationary-phase approximation (4) to $Z(m)$.
Or rather it is easy to compute its exponential factor, 
which is as far as we will carry our analysis.
One obtains at the saddle-point $\gamma$,
$S = - \int dx \sqrt{-2 m W(x)}$
or equivalently 
$$ 
         iS = -i \int dx \sqrt{-2 m W(x)}  \ ,
$$
where we of course take 
the positive square root of the positive number $-2mW(x)$.
When we carry $m$ back through the upper half-plane 
to the positive real axis, 
$\sqrt{-m}$ arrives as $-i\sqrt{m}$, 
whence $iS$ becomes, finally
$$
            iS = - \int\limits_{x_0}^{x_1} dx \sqrt{2 m W(x)}  \ .
$$
The exponential $\exp(iS)$ then yields the familiar WKB damping factor
$$
          \exp \left(- \int_{x_0}^{x_1} dx \sqrt{2 m W(x)}\right) \ .
          \eqno(12)
$$
Here we have obtained it via analytic continuation 
from a saddle-point approximation to 
the real-time path-integral with negative mass-parameter $m$.

Notice, incidentally, that we have all along been assuming that $W(x)<0$
at both the ``source'' $x_0$ and ``sink'' $x_1$.  
In the contrary case
where at least one of $x_0$ and $x_1$ is in the allowed region a 
global rotation of the sign of $m$ 
(though it would still be a valid operation)
would no longer secure us a saddle-point
trajectory valid for all relevant $x$.  
It would be logical in that case to make $m$ be a function of
position and rotate its sign only in the forbidden region.  We would
then obtain a hybrid formula approximating (7) as the product of
(12) from the forbidden region with the corresponding ordinary
semiclassical phase-factor from the allowed region.

As conducted above, our derivation had nothing to do with imaginary time
since the analytic continuation we employed touched only the
mass-parameter $m$, leaving $dt$ entirely alone.  However, the idea of
analytically continuing $dt$ to a complex variable does arise naturally
if one organizes in a certain order the infinite number of implicit
integrations that a path-integral contains.
Notice first that in (7) 
with $S$ given by (10), 
the expression $d\gamma$ 
can be construed as an infinite product of factors $d(dx)$ and $d(dt)$,
one for each ``infinitesimal segment'' of $\gamma$.\footnote{$^\star$}
{Equivalently, but with a symbolism that might look less confusing, we
 could say with reference to (9) that $d\gamma$ can be
 construed as an infinite product of factors $d\xdot$ and $d\tdot$.}
Observe moreover that 
the distinct increments $dt$ 
(though not $dx$) 
will act as independent integration variables
in our constant-energy path-integral, since it
puts no constraint on their sum $t_1-t_0$.
If now we reserve the integrations over the $dx$ for last, 
our path-integral resolves itself initially into a product of
independent one-dimensional integrals of the form
$$
    \int\limits_{dt=0}^\infty d(dt) e^{i ({m\over 2}{dx^2\over dt}-W(x)dt)}
    \ ,
    \eqno(13)
$$
where the limits of integration express the fact that the
nonrelativistic worldline $\gamma$ travels monotonically forward in
time.

But we've seen this type of integral before!  
It's essentially the integral we used 
to illustrate the saddle-point method, 
and as we saw then, 
it can in some sense be evaluated exactly, 
depending however on the unknown value of the prefactor or ``measure factor''.
More cautiously, 
we can observe that if we complexify $dt$
then the integrand of (13) 
has a pair of complex saddle-points
at $dt=\pm idx \sqrt{m/2W(x)}$.  
Of these, the saddle point on
the negative imaginary axis is the only one which is accessible
in the sense in which we used that word earlier.
That is, 
our original integration contour $(0,\infty)$
can be deformed to $-i(0,\infty)$
without changing the value of the integral.
The result is the Wick rotated integral
$$
      -i \int\limits_{d\tau=0}^{\infty} d(d\tau) 
     e^{{-m\over 2}{dx^2\over d\tau}-W(x)d\tau}     \eqno(14)
$$

About this Wick rotated integral we can make several comments. 
Firstly,
although it was motivated by 
the existence of a saddle point for imaginary $dt$, 
the integral in itself is still exact.  
As such, 
it and its analogs in quantum field theory offer
convenient starting points for Monte Carlo simulations, 
since the
integrand is now real and positive (and formally convergent).
On the other hand, one must remember that
the contour-deformation that led to (14) is valid
only for $x$ in the forbidden region\footnote{$^\dagger$}
{Notice in this connection that the existence of a forbidden region is
 bound up with our having chosen to do the path-integral at constant
 energy $E$.  With initial and final times held fixed instead, there is
 no forbidden region, as long as $V(x)$ is bounded below.}
where $W(x)>0$.
In the allowed region we thus should keep $dt$ real.  
The corresponding saddle-point or
``semiclassical path'' will then have a real tangent in the allowed
region and an imaginary tangent in the forbidden region.

A second comment is that we can think of the Wick-rotated path-integral
in a somewhat different way that relates it more closely to our analytic
continuation in the mass (see the appendix of [2]).  Namely, we
can introduce a complex parameter $\zeta$ into the action-integral
(9) by replacing $\tdot$ everywhere with
$\zeta\tdot$.  When we rotate $\zeta$ from real to imaginary, the
relative sign of the two terms is reversed and $iS/\hbar$ itself goes from
pure imaginary to pure real, the result being essentially the same as
(14).  Viewed in this manner, a Wick rotation appears
as a combination of two analytic continuations, one changing the sign of
the mass and the other changing $\hbar$ from real to imaginary.

As illustrated by these examples, the use of imaginary time is only one
device among many, that sometimes helps to simplify the approximate (or
exact) evaluation of the path-integral.
It has no independent status of its own and in general is neither the
beginning nor the end of the story.
We could for example 
plug the saddle-point approximation to (14) 
back into the full path-integral (7) 
to obtain the latter in the form of an integral 
over purely spatial paths,
corresponding to the so-called Jacobi action principle.  
If the measure-factor cooperated we could even render this
``Jacobian'' path-integral exact by evaluating (14) exactly.  
If not, 
it would represent a kind of cross 
between an exact path-integral and a saddle-point approximation to one.  
In either case, 
a further application of the saddle-point method to it, 
would immediately lead back to (12).

\section{Tunneling in quantum cosmology}
When it comes to cosmology, one must be cautious with the word
``tunneling'' since it now refers to the ``birth of a cosmos from
nothing''.  What's more, the relevant path-integral now involves a sum
over 4-geometries, which seems much farther from mathematical (or even
physical) respectability than what we were considering earlier.  Still,
if we are willing to neglect a number of important complications, we can
make the problem seem remarkably close to more homely examples of
tunneling, like the one treated above.  This of course, is basically the
view people have taken in discussing ``creation of the universe via a
gravitational instanton''.

To simplify things as much as possible, and to make the analogy with
one-dimensional tunneling as close as possible, let us consider an
``empty'' and (spatially) spherical cosmos which is born with zero
radius and subsequently expands.  Assume further that the amplitude for
such a process can be expressed as an integral of the form (7),
where now $\gamma$ represents a spacetime geometry and $S(\gamma)$ is a
suitable gravitational action, including boundary contributions as
needed.  In line with the spirit of this paper, I will take $\gamma$ to
carry a metric which is almost everywhere of Lorentzian signature.  If
we think of its initial portion as having the shape of a cone, then the
metric cannot actually be Lorentzian 
at the ``tip'' or {\it origin},\footnote{$^\flat$}
{Truly an origin here --- not only of the coordinates, but of the cosmos.}
but it can still be globally smooth if it is a ``Morse metric'' of the
type discussed in [3].  For definiteness, let us imagine the
domain of the gravitational path-integral to consist of metrics of this
type, in which case we can neglect any extra contribution to $S(\gamma)$
from the origin.  (In 1+1 dimensions there would be such a contribution,
as also in 3+1 dimensions for other topologies.  See [4] and
[5] for extensive discussion of this question.)

In writing down the action-integral $S$ explicitly, we could of course
employ proper time $\tau$ and spatial radius $a$ as our coordinates, but we
can lend $S$ a much more convenient form for present purposes by
employing instead the quantities $T$ and $v$, where on any hypersurface
$\Sigma$ coinciding with a 3-sphere of homogeneity, 
$v$ is the volume of $\Sigma$ ($v=2\pi^2 a^3$) 
and $T$ is the spacetime volume which has accumulated up to that stage,
or in other words the total 4-volume to the past of $\Sigma$ ($dT=vd\tau$).  
(Thus, $T$ is the time-coordinate natural to unimodular gravity.)  
In these coordinates the gravitational action $S$ takes the form
$$
         S = \int \; 
	     {-1\over 3 \kappa} \, {dv^2\over dT} 
	     + {k\over 2\kappa}\, v^{-2/3} dT
             - \Lambda \, dT
           \ ,                      \eqno(15)
$$
where $\kappa=8\pi G$,
$k=3 \cdot 2^{5/3} \pi^{4/3}\approx43.8,$\footnote{$^\star$}
{A 3-sphere of volume $v$ has Ricci scalar-curvature $R=kv^{-2/3}$}
and we assume $\Lambda>0$.
Henceforth, I will set $\kappa=1$.

Comparing (15) with (8) or (10) we can
recognize an analogy that makes
$T$ correspond to time and $v$ to position, 
while (up to sign)
the cosmological constant $\Lambda$ corresponds to the energy $E$.  
If we now imagine a ``source'' at the origin 
(representing perhaps the end of the previous cosmic cycle) 
and a ``sink'' at some other value of $v$, 
we have a nearly perfect analog of our tunneling problem.
The only thing spoiling the analogy is the sign of the kinetic term in
(15), which is negative\footnote{$^\dagger$}
{This, of course, is an instance of the ``wrong sign of conformal
 modes'' that invalidates the usual type of Wick-rotation for the
 gravitational path-integral.  The latter, incidentally, deforms the
 metric into the complexes such that $\Imag(g_{ab})>0$,
 cf. [4].} 
as if the ``mass'' were $-2/3\kappa$.
If we were not restricting ourselves to spherical symmetry, 
this minus sign in the direction of isotropic expansion 
would render the overall sign of the ``kinetic energy'' indefinite, 
and this in turn would complicate any attempt to discover a valid
saddle-point approximation to the path-integral.  
However, in our simplified situation, it suffices just to reverse the
overall sign of the action $S$.  
Since this merely converts the integrand $\exp(iS)$ into its complex conjugate, 
it clearly will have the same effect on the ``partition function'' $Z$
(assuming the ``measure factor'' in $d\gamma$ to be real).
In setting up our analogy with tunneling, then,
we can take $m=2/3$, $t=T$,  $E=\Lambda$, and $V=k/2v^{2/3}$.
(Notice also that  neither 4-volume here nor elapsed time there is
to be held fixed in (7), the analogy being perfect in that sense
as well.)

With these substitutions, our cosmological path-integral 
becomes a special case of the non-relativistic path-integral
which we evaluated in the previous section,
or rather which we approximated by various methods.  
Since the answer we obtained there, namely (12), 
was real to the accuracy to which we were working, 
undoing the complex conjugation doesn't change it, and we are left with
$Z\approx\exp(-I)$ 
as the saddle-point approximation to our path-integral,
where $I$ is the integral in (12) with the above substitutions.  
Let us also take for $v_1$ the special radius (or rather volume)
at which the cosmos can continue its expansion classically 
as a de Sitter spacetime.
We can obtain this radius from the  ``turning  point equation'',
$V=E$ or $\Lambda=k/2v_1^{2/3}$.
The integral then yields the simple answer\footnote{$^\flat$}
{One might worry that our gravitational potential $V(v)$ is infinite at
 $v=0$, but the corresponding divergence in the integrand of $I$ is
 only $1/v^{1/3}$.}
$$
    I = \int_0^{v_1} dv \sqrt{2m(E-V)} = {12 \pi^2 \over \Lambda}  
$$
and the consequent approximation
$$
       Z \approx \exp(- {12 \pi^2 \over \Lambda})  \ . \eqno(16)
$$

Notice that this amplitude, if treated as the square root of a
probability, would favor large values of $\Lambda$ over small ones, a
result that is entirely logical if one observes that small $\Lambda$
corresponds to big radius at the turning point, meaning that the cosmos
has to tunnel much farther during its ``quantum era''.  Notice also 
that (16) expresses damping rather than amplification for a
very good reason.  It represents an integral over a space of Lorentzian
metrics, none of which is a point of stationary phase.  That in this
situation the resulting cancellations lead to an exponentially small
answer is just what one would expect to happen.  From this point of
view, arguments that have tried to associate a small $\Lambda$ with an
exponentially large amplitude [6] [7]
seem to be guilty of working with the wrong saddle-point (``instanton'')
of the analytically continued action.
In effect, they have overlooked the conditions of validity that govern
which saddle-points are accessible and which are not.\footnote{$^\star$} 
{For the problem at hand, and for $\Lambda$ and $v_{final}$
 fixed at real values, there seems to be a total of six such
 saddle-points, if I haven't miscounted.}

Of course any conclusion like the above remains tentative, because no
one has ever (to my knowledge) embedded amplitudes like (16) in
a coherent theory of quantum cosmology.  One can compute some
approximate amplitudes, but one doesn't know what they mean.  
Should one construe them as values of  ``the wave function of the
universe'', or (as I think) would it be better to conceive of them as
contributions to the gravitational 
``quantal measure'' [8],
a quantity which in turn might find its ultimate interpretation in terms
of notions like
``preclusion'' [9]
and 
``anhomomorphic logic'' [10].

Into these questions intrude other conceptual issues,
which however are also more technical in nature.  
For example, 
how literally should one take the identification of our parameter $T$ 
as a kind of time?  
To embrace that idea is to do unimodular gravity, 
and thence spring other questions 
concerning for example 
selfadjointness of the unimodular Hamiltonian (see e.g. [5]).  
Which selfadjoint extension should we choose, 
or, instead of selecting one, 
should we discard unitarity altogether, 
in order to
provide for transitions to other topologies or to non-geometric phases?
In path integral terms: How should we treat topology changing histories?
And then one needs to confront the effects of anisotropy and inhomogeneity.
It can happen, for example, that a saddle point which might have
seemed accessible within a smaller space of highly symmetrical metrics
becomes inaccessible when the constraint of symmetry is relaxed
(compare e.g. the discussion of Taub vs. Friedmann in [5]).

Rather than continuing to multiply questions in this vein, let me close
this section with a thought about the possible 
``practical'' meaning of the damping-factor we have computed.
Given the rather fanciful assumption of a spherical and empty cosmos, we 
might attach to $Z$ in equation (16)
a set of words like
``the amplitude for the cosmos to arrive at size $v$ starting from zero size''.
This phrase elicits a picture in which the cosmos is ``born under a
potential barrier'' and then develops through different phases of
growth.  In the earliest phase, it likely has no continuum description
at all.  (Perhaps it has a tree-like structure [11].)
Later it begins to expand as a Lorentzian manifold, but, being empty,
must do so in a quantal manner by ``tunneling''.  Finally, it emerges as
a classical universe with a size dictated by the size of $\Lambda$,
which 
(despite causal set arguments that $\Lambda$ must fluctuate) 
we have taken 
to be a fixed parameter.  
A question then is how long the cosmos
spends in the first two phases.  
For this we can return to our nonrelativistic
example where (12) governs the tunneling rate per unit time.  
Treating $T$ here as our time parameter, we might thus think of
(16) (or its square) as a rate of tunneling per birth of causal
set element (since 4-volume simply measures number of elements).  
On this reading, 
our computation of $Z$ would in effect have been counting 
the average number of births taking place during the pre-geometric
and tunneling phases of cosmic growth.







\section{Summary}
Probably few people believe that clocks begin to tick off imaginary
seconds the moment some alpha-particle finds itself under a
potential barrier.  And yet in the context of the gravitational
path-integral, something like this notion has often been put forward.
In opposition to such notions, 
a more prosaic (and more traditional) viewpoint might maintain
that a tunneling amplitude, 
like all other quantal amplitudes, 
is given in the first place by a real-time path-integral.\footnote{$^\dagger$}
{The analytically continued metrics used in connection with black hole
 thermodynamics present a somewhat different case than tunneling.  
 An imaginary time-parameter arises because the operator of interest in
 equilibrium statistical mechanics is $\exp(-\beta\hat{H})$ rather than 
 $\exp(-i\hat{H}t)$.  One is computing a sum over stationary states, not
 dealing with a process developing in time.}
In previous sections, I have tried to illustrate a few of the different
tools that are available for approximating tunneling amplitudes, 
and also to illustrate that
some of them (like analytic continuation in the mass)
are simpler in conception than those (like Wick rotation) 
which need to introduce a space of complexified paths $\gamma$. 
Thus, we considered several integrals and several different ways of
treating them.

To begin with,
we approximated the simple integral (2), 
in two different ways,
first by
deforming the integration contour into the complexes, and second by
complexifying the real parameter $A$ without modifying the domain of
integration.  
With the first approach, we had to pay heed to several
conditions of validity: 
the integrand had to continue to an analytic function;
the complexified contour $\Gamma$ had to be deformable back to the
original contour without changing the value of the integral;
$\Gamma$ had to be sufficiently close to either a steepest-descent or a
stationary-phase path;
and the exponent $f(z)$ had to be ``sufficiently rapidly varying'' in
the relevant sense.
These conditions involved the behavior of the analytically continued
integrand at infinity, and more generally some understanding of its
overall ``topography''.
Ultimately, they picked out a unique saddle point
and a unique sign for the consequent approximation (4).  
With the second approach, related conditions of validity determined the
direction of continuation of the parameter $A$ and governed the validity
of the resulting approximation.

We then turned to a path-integral for quantal tunneling, viewed as an
infinite dimensional analog of the previous problem.  Similar approaches
were feasible and similar conditions of validity applied, the main
difference being that those conditions were much more formal and much less
clearly complete than in the one-dimensional case.  
Our first treatment of tunneling rested on complexification of the
particle-mass $m$, which we continued from the positive to the negative
real axis, in order to convert our integral $Z(m)$ into one to which the method
of stationary phase would apply.  We then continued the answer back to
positive $m$ and recognized our result as the familiar damping factor (12).
The fact that we obtained damping rather than amplification
was not arbitrary, but followed from the direction in which we continued 
$m$, and this in turn was dictated by 
the condition that the integral $Z(m)$ continue to be defined
for the intermediate values of $m$.
Our second treatment of the tunneling problem complexified $dt$ instead
of $m$.  
This led to a hybrid path-integral that was ``Wick rotated'' in the
forbidden region, but not in the allowed region.\footnote{$^\flat$}
{In contrast the first method could get by without treating the two
 regions differently, as long as both $x_0$ and $x_1$ belonged to the
 forbidden region.}
A steepest descent approximation to this integral then reproduced the
same damping factor (12).  Along the way, we obtained a purely
spatial path-integral corresponding to the so-called Jacobi action
principle.
At the risk of repetition, let me emphasize that with both methods, the
fact of damping as opposed to amplification is seen to be a property of
the real-time path-integral itself.  It emerges automatically, when one
takes into account the need for consistency in the approximation method.

Finally, we carried some of this analysis over to the problem of the
birth of a cosmos in quantum cosmology.  Leaving to one side the thorny
conceptual and technical problems that arise in this context, we wrote
down a path integral describing a process in which a cosmos is born as a
sphere of zero radius and expands quantum mechanically (in a ``tunneling
regime'') until it is big enough to continue its expansion as an
approximately classical (de Sitter) spacetime.  This sum over Lorentzian
geometries produced a problem identical in form to our earlier example
of tunneling in nonrelativistic quantum mechanics.  We were therefore
able to solve it easily (at leading semiclassical order), and to
conclude that the answer takes the form of a WKB damping factor.  To the
extent that the original integral is appropriate to the physics, this
conclusion seems to be inescapable.  In particular, it is free of any
ambiguity arising from the need to choose among alternative
saddle-points during the approximation process.

More generally in quantum gravity, one might expect to be able to rule
certain saddle-point metrics out or in if one could pin down more of the
conditions of validity that pertain to the particular situation at hand.
For example reference [5] exhibited a number of saddle-point
metrics that arise within unimodular quantum cosmology and considered
different criteria for accepting or rejecting them.  It seems that an
analysis like that in the previous section could clarify further
the status of some of those complex metrics and their relatives.





\bigskip
\noindent
Thanks go to David Rideout for his help with the figures 
and to Adam Brown for correcting a wrong equation (6) in a
previous version of this paper.
Research at Perimeter Institute for Theoretical Physics is supported in
part by the Government of Canada through NSERC and by the Province of
Ontario through MRI.

\ReferencesBegin

\ref [1] 
Jon Mathews and R. L. Walker,
{\it Mathematical Methods of Physics}
(Second Edition)
(Addison-Wesley 1970)

\ref [2] 		
  Rafael D.~Sorkin,
``Forks in the Road, on the Way to Quantum Gravity'', talk 
   given at the conference entitled ``Directions in General Relativity'',
   held at College Park, Maryland, May, 1993,
   published in
   \journaldata{Int. J. Th. Phys.}{36}{2759--2781}{1997},
   \eprint{gr-qc/9706002},
   \hpf{ http://www.physics.syr.edu/~sorkin/some.papers/ }

\ref [3] 
A.~Borde, H.F.~Dowker, R.S.~Garcia, R.D.~Sorkin and S.~Surya,
``Causal Continuity in degenerate spacetimes'',
 \journaldata {Class. Quant. Grav.}{16}{3457-3481}{1999}
 \eprint{gr-qc/9901063}

\ref [4] 
 Jorma Louko and Rafael D Sorkin,
``Complex Actions in two-dimensional topology change'',
  \journaldata{Class. Quant. Grav.}{14}{179-203}{1997}
  \eprint{gr-qc/9511023}

\ref [5]
 Alan Daughton, Jorma Louko and Rafael D.~Sorkin,
``Instantons and unitarity in quantum cosmology with fixed four-volume'',
 \journaldata {Phys. Rev.~D} {58} {084008} {1998} 
 \eprint{gr-qc/9805101}	

\ref [6]
S.W.~Hawking, ``The cosmological constant is probably zero'',
\journaldata{Physics Letters}{B134}{403-404}{1984}

\ref [7]
Sidney Coleman, ``Why there is nothing rather than something: a theory
 of the cosmological constant'',
 \journaldata{Nuclear Physics}{B 310}{643-668}{1988}

\ref [8] 
Rafael D.~Sorkin,
``Quantum Mechanics as Quantum Measure Theory'',
   \journaldata{Mod. Phys. Lett.~A}{9 {\rm (No.~33)}}{3119-3127}{1994}
   \eprint{gr-qc/9401003}

\ref [9]   
Robert Geroch, ``The Everett interpretation", 
\journaldata{No\^{u}s}{18}{617-633}{1984};
\sepref
  Rafael D.~Sorkin,
``Quantum Measure Theory and its Interpretation'', 
  in
   {\it Quantum Classical Correspondence:  Proceedings of the $4^{\rm th}$ 
    Drexel Symposium on Quantum Nonintegrability},
     held Philadelphia, September 8-11, 1994,
    edited by D.H.~Feng and B-L~Hu, 
    pages 229--251
    (International Press, Cambridge Mass. 1997)
    \eprint{gr-qc/9507057}
  \hpf{http://www.physics.syr.edu/~sorkin/some.papers/}

\ref [10]  
Rafael D. Sorkin,
``Quantum dynamics without the wave function''
 \journaldata{J. Phys. A: Math. Theor.}{40}{3207-3221}{2007}
 (http://stacks.iop.org/1751-8121/40/3207)
\eprint{quant-ph/0610204} 
\hpf{http://www.physics.syr.edu/~sorkin/some.papers/}

\ref [11]
Rafael D.~Sorkin,
``Indications of causal set cosmology'',
 \journaldata {Int. J. Theor. Ph.} {39 {\rm(7)}} {1731-1736} {2000}
 (an issue devoted to the proceedings of the Peyresq IV conference,
  held June-July 1999, Peyresq France),
 \eprint{gr-qc/0003043},\lbr
 \hpf{ http://www.physics.syr.edu/~sorkin/some.papers/ }

\ref [12] 
 Rafael D.~Sorkin, 
``On the Role of Time in the Sum-over-histories Framework for Gravity'',
    paper presented to the conference on 
    The History of Modern Gauge Theories, 
    held Logan, Utah, July 1987, 
    published in  
    \journaldata {Int. J. Theor. Phys.}{33}{523-534}{1994}


\pagebreak

 \def\FigureCaption#1#2{\vbox{
 \leftskip=1.5truecm\rightskip=1.5truecm     
 \singlespace
 \noindent{\it Figure #1}. #2
 \vskip .25in\leftskip=0truecm\rightskip=0truecm}
 \sesquispace}


\epsfbox{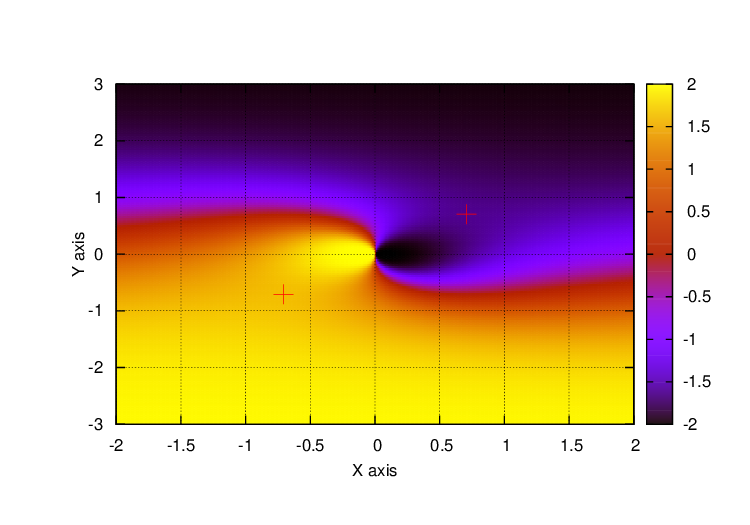}

\FigureCaption{1}
{The colors depict the real part of the exponent $f(z)$,
 the range $(-\infty,\infty )$ having been compressed into $(-2,2)$.
 The saddle points $z_\pm$ are marked with plus signs. }

\bigskip






\end                                         


(prog1    'now-outlining
  (setq outline-initial-level 8)          
  (Outline 
     "\f......"
      "
      "
      "
   ;; "\\\\message"
   "\\\\Abstrac"
   "\\\\section"
   "\\\\subsectio"
   "\\\\appendi"
   "\\\\Referen"
   "\\\\ref....[^|]"
  ;"\\\\ref....."
   "\\\\end